\documentclass[%
twocolumn,
showpacs,
bibnotes,
amsmath,amssymb,
aps,
prb,
]{revtex4-1}

\usepackage{graphicx}
\usepackage{dcolumn}
\usepackage{bm}

\usepackage{ulem} 
\usepackage{color}
\usepackage{multirow}
\newcommand{\ra}[1]{\renewcommand{\arraystretch}{#1}}
\usepackage{array,ragged2e}
\usepackage{booktabs}

\begin{document}


\title{Lattice-matched heterojunctions between topological and normal
  insulators: A first-principles study}

\author{Hyungjun Lee}
\email{hyungjun.lee@epfl.ch}
\affiliation{Institute of Physics, \'{E}cole
  Polytechnique F\'{e}d\'{e}rale de Lausanne (EPFL), CH-1015 Lausanne,
  Switzerland}
\author{Oleg V. Yazyev}
\affiliation{%
  Institute of Physics, \'{E}cole
  Polytechnique F\'{e}d\'{e}rale de Lausanne (EPFL), CH-1015 Lausanne,
  Switzerland}%

\date{\today}

\begin{abstract}
  Gapless boundary modes at the interface between topologically
  distinct regions are one of the most salient manifestations of
  topology in physics.
  Metallic boundary states of time-reversal-invariant topological
  insulators (TIs), a realization of topological order in condensed
  matter, have been of much interest not only due to 
  such a fundamental nature, but also due to their practical
  significance. These boundary states are immune to backscattering and
  localization owing to their topological
  origin, thereby opening up the possibility to tailor them for
  potential uses in spintronics and quantum computing. The
  heterojunction between a TI and a normal insulator (NI) is a representative playground for exploring such a topologically protected metallic boundary state and expected to constitute a building
  block for future electronic and spintronic solid-state devices based on TIs.
  Here, we report a
  first-principles study of two experimentally realized lattice-matched heterojunctions
  between TIs and NIs, Bi$_2$Se$_3$(0001)/InP(111) and Bi$_2$Te$_3$(0001)/BaF$_2$(111).
  We evaluate the band offsets at these interfaces
  from many-body perturbation theory within the $GW$ approximation as well as density-functional theory.
  Furthermore, we investigate the topological interface states, demonstrating that at these lattice-matched heterointerfaces they are
  strictly localized and their helical spin textures are as well preserved as those at the
  vacuum-facing surfaces. These results taken together may help in
  designing devices relying on spin-helical metallic boundary states of TIs.
\end{abstract}


\maketitle


\section{\label{sec:1}Introduction}

Interfaces have been a fertile ground for creating novel states and
exploring exotic physics in the history of
condensed-matter physics due to a myriad of intriguing phenomena
emerging at them
\cite{Ohtomo2004,Okamoto2004,Reyren2007,Caviglia2008,Gozar2008,Valencia2011,Hwang2012,Chakhalian2014}. 
In addition to such a fundamental interest in
  interfaces, they have attracted considerable attention also
from an applications standpoint. In modern solid-state devices
permeating our daily life, interface formation is inevitable due
to their inherent heterojunction structures, and thus thorough
understanding of interface-related phenomena is indispensable for
manipulating their functionality. One of the archetypal examples is the metal-oxide-semiconductor field-effect
transistor (MOSFET), the workhorse of modern
microelectronics\cite{Sze2007}. In this regard, there have been much
theoretical and experimental interest in a variety of phenomena occurring at the interface of heterojunctions such as band offsets and
Schottky barriers\cite{Tersoff1984,Peressi1998,Robertson2013}.

Rich interface physics arises also in the context of
topology in condensed matter in the sense that gapless boundary states,
one of the key emergent topological phenomena, manifest themselves at the
interface that separates topologically distinct phases\cite{Altland2010,Fradkin2013}. These boundary states are
ensured by the different topologies of the constituent bulk
phases\cite{Altland2010,Fradkin2013}, referred to as the bulk-boundary correspondence\cite{Franz2013}, and ubiquitous
in various contexts of physics\cite{Eschrig2011,Wilczek2016}.
Initial research in condensed-matter physics along this line
includes solitons in
polyacetylene by Su, Schrieffer, and Heeger\cite{Su1979}, which is a
condensed-matter realization of the
earlier Jackiw-Rebbi model in high energy physics\cite{Jackiw1976}, and the integer quantum Hall effect\cite{Klitzing1980,Thouless1982}.
More recently, research on topology in condensed matter has been
rekindled by the discovery of time-reversal-invariant topological
insulators (TIs) including the quantum spin Hall systems\cite{Hasan2010,Qi2011,Bernevig2013}.

Since their first theoretical proposals\cite{Kane2005Z2,Kane2005Graphene,Bernevig2006}, interest in TI has surged among both science and engineering
communities owing to the fact that its topological phenomena don't
require extreme conditions such as low temperatures and high external
magnetic fields and not only it might serve as a route to realizing Majorana
fermions and magnetic monopoles, but also its topologically protected
spin-helical metallic boundary states might pave the way for future
spintronics and quantum computing\cite{Hasan2010,Qi2011,Moore2010}.
While initial research efforts had mostly been devoted to the vacuum-facing TI
surfaces\cite{Hsieh2008,Hsieh2009Science,Hsieh2009Nature,Chen2009,Yazyev2010,Park2010,Moon2011},
increasing experimental\cite{Liu2013,Berntsen2013,Yoshimi2014,Landolt2014} and theoretical attention\cite{Wu2013,Menshov2015,Seixas2015,Chen2015,Kufner2016}
have recently been paid to the more realistic situation, the interfaces between
TIs and normal insulators (NIs). 
It is motivated by the fact that interfaces are
  protected from the possible ambient
contamination\cite{Kong2011,Kim2014} and moreover these types of heterojunctions can be integrated into existing
semiconductor technology, hence, they are more advantageous for
utilizing the topological conducting boundary states.

Among TIs realized experimentally to date, the
Bi$_2$Se$_3$ family compounds are prototypical on account of their simple
surface-state band dispersion with the single Dirac cone and a
relatively large bulk band gap of greater
than 0.1~eV\cite{Hsieh2008,Hsieh2009Science,Hsieh2009Nature,Chen2009}. They are currently synthesized by using various
methods such as chemical vapor deposition, Bridgman growth, and
molecular beam epitaxy (MBE), and among them the MBE technique is
preferable to other approaches in that it naturally allows for
heterojunctions with potential solid-state-device
applications\cite{He2013}. Although it is believed that due to the van
der Waals epitaxy characteristic of this class of compounds, 
the lattice match between TI deposit and NI substrate is not a critical factor\cite{He2013}, small lattice mismatch is still responsible for
high-quality of TI films \cite{Tarakina2012,Guo2013,Schreyeck2013,Caha2013} yielding, in particular, good
transport properties\cite{Hoefer2014}.

In the present work, we consider the experimentally realized
lattice-matched TI/NI heterojunctions, Bi$_2$Se$_3$(0001)/InP(111) with the lattice
misfit of 0.2\,\%\cite{Tarakina2012,Guo2013,Schreyeck2013}
and Bi$_2$Te$_3$(0001)/BaF$_2$(111) with that of
0.1\,\%\cite{Caha2013,Hoefer2014}. For these heterojunctions, we
obtain the band offsets at the interfaces via the quasiparticle $GW$ approximation as
well as semilocal density-functional theory and investigate the
electronic and spin structures of topological interface states. From the calculations, we demonstrate
that at these lattice-matched heterointerfaces topologically protected
interface states are strictly localized
and their helical spin textures are as well maintained as those at the
vacuum-facing surfaces.  These results taken together may help in
designing future spintronic and electronic devices utilizing the
topologically protected spin-helical metallic boundary states of TIs.

The remainder of this paper is organized as follows: In
Sec.~\ref{sec:2}, we describe computational details employed in this study.
Sec.~\ref{sec:3} discusses the main results of our work. Namely, in
Sec.~\ref{sec:3a} we introduce the interface models adopted here and in Sec.~\ref{sec:3b} we describe the band offsets in
heterojunctions considered and the electronic and spin structures of topological interface states.
Finally, Sec.~\ref{sec:4} concludes our paper.

\section{\label{sec:2}Computational Method}

Our present study is based on \textit{ab initio}
density-functional-theory (DFT) method\cite{Hohenberg1964,Kohn1965} as
implemented in the {\sc Quantum ESPRESSO} package\cite{Giannozzi2009}
and many-body perturbation theory within the Hedin's $GW$ approximation\cite{Hedin1965} as implemented in the {\sc Yambo} code\cite{Marini2009}.
In DFT calculations, the generalized gradient approximation of Perdew-Burke-Ernzerhof (PBE) type\cite{Perdew1996} is
employed for the exchange-correlation energy and the
norm-conserving pseudopotentials (PPs)\cite{Hamann1979} with multiple
projectors per orbital angular momentum channel\cite{Hamann2013} are
used to simulate the interaction of valence electrons with atomic
cores.
Spin-orbit coupling is treated by the fully-relativistic PPs\cite{Kleinman1980} in fully
separable forms\cite{Corso2005}.
Wave functions are expanded in terms of plane waves with the kinetic energy
cutoffs of 120~Ry for Bi$_2$X$_3$ (X=Se, Te) and 180~Ry for
InP, BaF$_2$, and all interface models. The $\bm{k}$-point meshes of 12$\times$12$\times$1 and
12$\times$12$\times$12 in the scheme of Monkhorst-Pack\cite{Monkhorst1976} are used for the
Brillouin-zone sampling in interface and
bulk calculations, respectively.
These computational parameters were carefully checked and chosen to
allow for a numerical precision of better than 0.1~meV for the total
energy per atom in each case.

For the purpose of investigating the band alignment in the heterojunction between bulk materials $A$ and $B$ ($A/B$), the
following expression is used\cite{Walle1987,Baldereschi1988}: 
\begin{align}
  E_{\text{v(c)},A/B}^\text{BO} &= \Delta E_{\text{VBM(CBM)},A/B}+\Delta V_{A/B}\nonumber\\
                           &= E_{\text{VBM(CBM)},A}-E_{\text{VBM(CBM)},B}+\Delta V_{A/B}
\end{align}
for the valence (conduction)-band offset (BO), where
$E_{\text{VBM},X}$ and $E_{\text{CBM},X}$ are, respectively, the
valence-band maximum (VBM) and the conduction-band minimum (CBM) of a
bulk material $X$ ($X=A,B$), and $\Delta V_{A/B}$ is the potential lineup across
the interface from $B$ to $A$. The band-edge positions of $E_{\text{VBM},X}$ and $E_{\text{CBM},X}$ are calculated with respect to the reference of
each bulk constituent $X$ which is determined by the macroscopic average\cite{Baldereschi1988} of electrostatic potential comprising the local part of PPs and
the Hartree potential through bulk calculations. The potential lineup $\Delta V_{A/B}$
is obtained also by using the macroscopic average
method through interface
calculations. 
To be precise, in addition to DFT-PBE the band-edge position is obtained by the many-body perturbation
theory within the $GW$ approximation at the $G_0W_0$ level\cite{Hybertsen1986}, but the potential lineup at interface is
derived only from the DFT-PBE
calculations, following the previous
studies\cite{Shaltaf2008,Steiner2014}.
Finally, the band-edge positions are determined from Wannier
interpolation\cite{Yates2007} on a dense
$\bm{k}$ grid of 100$\times$100$\times$100 using the {\sc Wannier90} package\cite{Mostofi2014}.

\begin{figure}
  \centering
  \includegraphics[width=0.475\textwidth]{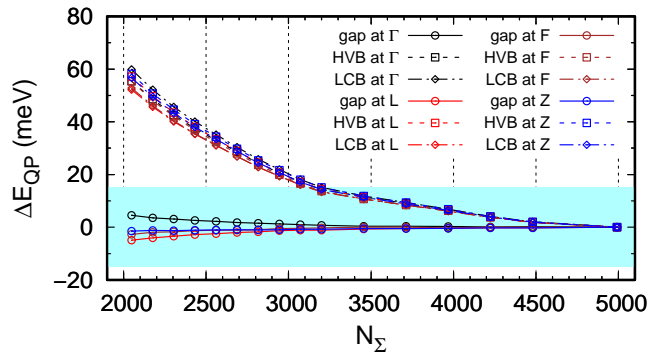}
  \caption{(Color online) Convergence study of quasiparticle (QP)
    energies $E_\text{QP}$ at high-symmetry $\bm{k}$ points of
    2$\times$2$\times$2 grid with respect to the
    number of bands $N_\Sigma$ in the summation of the correlation
    part of the self-energy for bulk
    Bi$_2$Se$_3$. Here, assuming that the kinetic energy cutoff of 60
    Ry for the polarizability and the number of bands of 4992 for the
    correlation self-energy summation are enough to reach the (numerically) converged QP energies, we plot the differences $\Delta E_\text{QP}$ from them as a function of $N_\Sigma$.
    The HVB (LCB) represents the energy of the highest valence (lowest
    conduction) band; the high-symmetry $\bm{k}$ points of $\Gamma$, L,
    F, and Z correspond to the $\bm{k}$ points of (0,0,0), (0.5,0,0),
    (0.5,0.5,0), and (0.5,0.5,0.5) in reciprocal lattice units,
    respectively. The cyan-shaded region indicates our convergence
    criterion of $\pm$15~meV for QP energies of bulk Bi$_2$Se$_3$.}
  \label{fig:Fig-conv}
\end{figure}

\begin{figure}[hb!]
  \centering
  \includegraphics[width=0.475\textwidth]{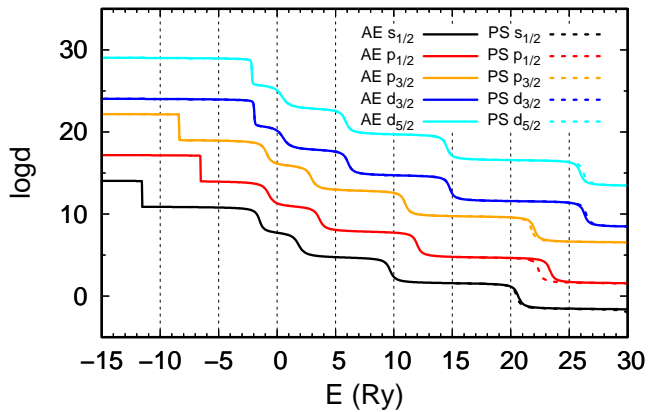}
  \caption{(Color online) Scattering properties of Bi atom. To
    describe scattering properties, the logarithmic derivative
      (logd) at
      $r=3.0~\text{Bohr}$ is calculated as a function of energy $E$ using the
      expression of $\arctan[r(d\psi_{l,j}/dr)/\psi_{l,j}]/\pi$, where $\psi_{l,j}$
      is an atomic radial wave function for the state with the orbital angular momentum $l$ and the total
      angular momentum $j$. AE and
      PS denote the all-electron and pseudo Bi atoms,
      respectively. All curves are offset for clarity.}
  \label{fig:Fig-logd}
\end{figure}

\begin{table}[t]
  \setlength{\extrarowheight}{3pt}
  \setlength{\tabcolsep}{5pt}
  \caption{Summary of convergence parameters for
      $G_0W_0$ calculations. $E_\chi$ stands for the kinetic energy
      cutoff in the plane-wave expansion of the polarizability $\chi$ and $N_\Sigma$ the number of both
      occupied and unoccupied states in the summation of the
      correlation part of the self-energy. In general, the
      number of bands required for convergence of the polarizability,
      $N_\chi$, is smaller than $N_\Sigma$, but for the sake of convenience we used the same value for
      $N_\chi$ as $N_\Sigma$.}
  \begin{tabular}{ccccccc}
    \hline
    \hline
    & $\bm{k}$ points & $E_\chi$ (Ry) & $N_\Sigma$,\,$N_\chi$\\
    \hline
    Bi$_2$Se$_3$ & $6\times 6\times 6$ & 48 & 3200\\
    Bi$_2$Te$_3$ & $6\times 6\times 6$ & 42 & 3200\\
    InP & $8\times 8\times 8$ & 50 & 2048\\
    BaF$_2$ & $8\times 8\times 8$ & 48 & 2048\\
    \hline
    \hline
  \end{tabular}
  \label{tab:Tab1}
\end{table}

In our $G_0W_0$ calculations, dielectric function is calculated within the
random phase approximation\cite{Adler1962,Wiser1963} and its frequency dependence is approximated by employing the plasmon-pole model of Godby and
Needs\cite{Godby1989}. In order to take fully into account the spin-orbit
coupling, two-spinor wave functions from the DFT-PBE
calculations are taken as inputs to the calculations of the
noninteracting Green's function and the screened Coulomb
interaction\cite{Sakuma2011}. All parameters relevant to
$G_0W_0$ calculations except for the number of
$\bm{k}$ points were determined in calculations with the smaller
$\bm{k}$-point mesh of 2$\times$2$\times$2 than that in actual
calculations as done in the previous
studies\cite{Malone2013,Klimes2014} and Figure~\ref{fig:Fig-conv}
illustrates the convergence behavior of
quasiparticle energies with respect to the number of states for the
correlation self-energy summation in bulk Bi$_2$Se$_3$. Finally, using
these obtained parameters we carried out additional convergence studies by
varying the number of $\bm{k}$-point grid.
Consequently, all parameters were chosen to
ensure the convergence of the energies of the highest valence and
lowest conduction states at the high-symmetry $\bm{k}$ points to within 15~meV except for the wide-gap insulator BaF$_2$ for
which the criterion is 50~meV. They are summarized in Table~\ref{tab:Tab1}.

Before leaving this section, it is important to mention that the special construction of PPs is essential
for obtaining more accurate results in $GW$
calculations\cite{Rohlfing1995,Marini2001,Shishkin2006,Umari2012,Malone2013,Steiner2014}. In order to describe
accurately the high-energy unoccupied states,
in particular necessary for $GW$
calculations due to their slow convergence behavior with respect to the
number of states\cite{Shih2010}, we constructed PPs so that their
scattering properties can be matched well to the all-electron
counterparts up to 20~Ry above the vacuum
level\cite{Shishkin2006,Steiner2014}. As an example, we show the plot
of the logarithmic
derivatives for Bi atom in Figure~\ref{fig:Fig-logd}. Additionally, for
In, Ba, and Bi atoms we constructed PPs with the whole atomic shell including semicore orbitals taken as valence so
as to correctly describe the exchange contribution to the self-energy, as suggested
by the previous literature\cite{Rohlfing1995,Marini2001,Umari2012,Malone2013}.

\section{\label{sec:3}Results and discussion}

\subsection{\label{sec:3a}Interface models}

We commence with the description of interface models adopted in
this study for two TI/NI heterojunctions. 
First, for the
Bi$_2$Se$_3$(0001)/InP(111) heterojunction we consider two heterointerface models
according to the topmost atomic layer of InP(111) substrate, Bi$_2$Se$_3$(0001) on
In-terminated InP(111) surface [hereafter InP(111)A] and on
P-terminated one [hereafter InP(111)B]. Experiments indeed reveal that
high-quality quintuple layers (QLs) of Bi$_2$Se$_3$ are epitaxially grown on both
InP(111)A and B surfaces by MBE\cite{Guo2013,Schreyeck2013}, whereas the previous theoretical study reported that 
the formation energy of Bi$_2$Se$_3$ on InP(111)A is lower by about
0.5 eV per ($1\times 1$) cell than that
on InP(111)B\cite{Guo2013}.
For each heterointerface model, we consider a total of six different
configurations consisting of combinations of the relative lateral position of
atoms (three possibilities) in the interfacial atomic layers and the relative orientation (two possibilities) between NI
substrate and TI deposit. Denoting the three in-plane lattice sites
allowed by symmetry as A, B, and C, the former can be expressed as $\cdots$CBA/ACCBB$\cdots$, $\cdots$ACB/ACCBB$\cdots$, and
$\cdots$BAC/ACCBB$\cdots$, and the latter as $\cdots$CBA/ACCBB$\cdots$ and $\cdots$BCA/ACCBB$\cdots$.
Due to the noncentrosymmetric bulk structure of InP, it is not possible to
construct the symmetric supercell for both of these two interface
models; instead, we construct
the asymmetric one in which two interfaces in supercell resulting from
the periodic boundary condition are modeled with the most and second-most
energetically favorable configurations determined as described later. 

Second, regarding the Bi$_2$Te$_3$(0001)/BaF$_2$(111) heterojunction, we consider
one interface model of Bi$_2$Te$_3$(0001) on F-terminated
BaF$_2$(111) surface and for this model, we examine six configurations as
in the case of the
Bi$_2$Se$_3$(0001)/InP(111) heterojunction. The centrosymmetry of bulk BaF$_2$ enables us to construct the symmetric supercell in this case.

For each interface model above, we obtain the corresponding supercell
structure without vacuum by determining the lowest-energy configuration and its optimized distance
between TI deposit and NI substrate.
To this end, we consider the simplified interface model: the structure composed of the 13-layer
InP(111) and 1-QL Bi$_2$Se$_3$ slabs with the vacuum of 20~\AA{}
for the Bi$_2$Se$_3$(0001)/InP(111) heterojunction and one composed of the 18-layer
BaF$_2$(111) and 1-QL Bi$_2$Te$_3$ slabs with the vacuum of the
same thickness for the Bi$_2$Te$_3$(0001)/BaF$_2$(111) heterojunction.
Starting from these minimal models with the in-plane lattice constants of
TIs (4.140~\AA\cite{Roy2014} and 4.386~\AA\cite{Nakajima1963}{} for
Bi$_2$Se$_3$ and Bi$_2$Te$_3$, respectively) set equal to those of NI
substrates (4.134~\AA\cite{Menoni1987} and 4.382~\AA\cite{Radtke1974}{}
for InP and BaF$_2$, respectively),
we obtained the optimized distance for each configuration by calculating the total energy as a function of the
distance between the blocks of Bi$_2$Se$_3$ (Bi$_2$Te$_3$)
and InP (BaF$_2$) and then fitting the obtained curve to the equation of
state\cite{Vinet1989}. Finally, we determined the lowest-energy
configuration for each model from comparisons among six possible configurations.

From these results, it is shown that in all simplified models, the
stacking sequence rotated by 60$^\circ$ across the
interface, i.e., $\cdots$BCA/ACCBB$\cdots$, gives rise to the configuration with the lowest energy. 
As for the relative lateral position and the optimized
distance, 
we found that the configuration with Se atoms on top of In atoms in the
interfacial layers
is energetically most favorable
for Bi$_2$Se$_3$(0001)/InP(111)A with the distance of
2.712~\AA{} and one with 
Se atoms on top of P atoms in the interfacial layers
is the
lowest-energy configuration for Bi$_2$Se$_3$(0001)/InP(111)B with the
distance of 2.594~\AA{}\footnote{%
In order to assess the influence of van der Waals correction,
  we carried out calculations using the semiempirical method by Grimme (PBE-D2)
[S. Grimme, J. Comput. Chem. \textbf{27}, 1787 (2006)]. While PBE-D2 leads to the certain decrease of the interatomic distances at
the interface of our heterojunction models,
we found that our main results obtained with DFT-PBE still hold.}.

For Bi$_2$Te$_3$(0001)/BaF$_2$(111), the
configuration with the same in-plane position of Te atoms in the
interface layer and Ba atoms in the first subinterface layer
is found to have the lowest energy with the distance
of 3.452~\AA{}. Using the obtained near-interface structures, in order to avoid the
spurious interaction between two opposite-side interfaces, we constructed
the supercell comprising 6-QL Bi$_2$Se$_3$ and 37-layer
InP(111)A or B, which amounts to about 120~\AA{} in length, for
Bi$_2$Se$_3$(0001)/InP(111)A or B and
one comprising 6-QL Bi$_2$Te$_3$ and 51-layer
BaF$_2$(111), amounting to about 125~\AA{}, for Bi$_2$Te$_3$(0001)/BaF$_2$(111).

\subsection{\label{sec:3b}Electronic structures of interfaces}

\subsubsection{\label{sec:3b1}Band alignment}

\begin{figure}
  \centering
  \includegraphics[width=0.475\textwidth]{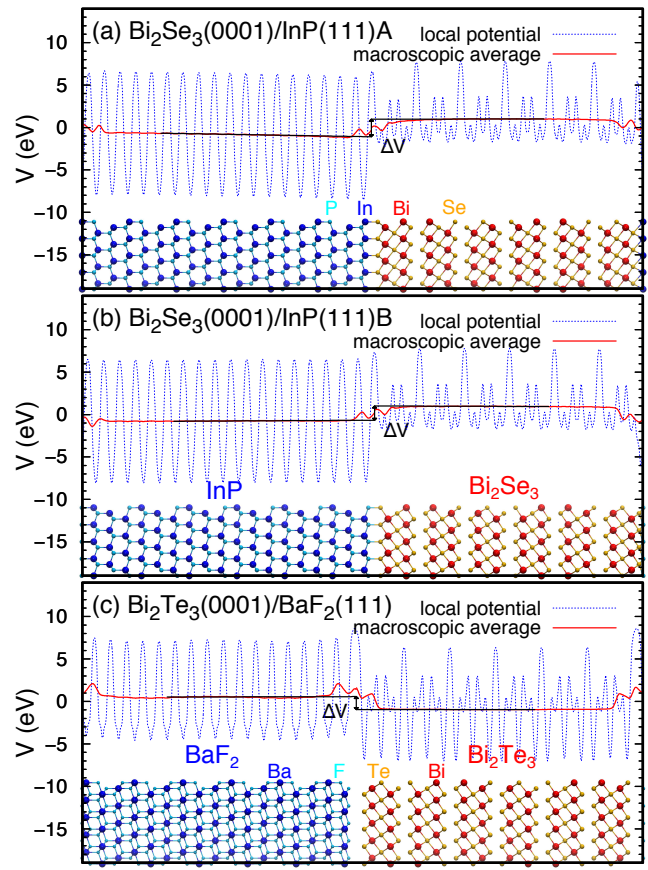}
  \caption{(Color online) Planar and macroscopic averages of electrostatic potential
    for (a) Bi$_2$Se$_3$(0001)/InP(111)A, (b)
    Bi$_2$Se$_3$(0001)/InP(111)B, and (c)
    Bi$_2$Te$_3$(0001)/BaF$_2$(111) heterojunctions. Blue dashed lines
    correspond to the in-plane averaged potential along the growth direction and red solid lines
    the corresponding macroscopic-averaged potential. Black solid lines denote the macroscopic-averaged potential in
    the bulk-like region of each interface component and the resulting potential
    offset $\Delta V$ at the interface is indicated by using the
    double-headed arrows and labels.
    Ball-and-stick models of heterojunction structures are given in the lower part of each panel with the
    atoms indicated with the corresponding labels and colors.}
  \label{fig:Fig-macro}
\end{figure}

\begin{table*}
  \caption{Summary of values related to the band alignment for
    heterojunctions considered. $E_\text{g}^\text{X}$ represents the
    fundamental band gap for the bulk material X, VBO (CBO) the
    valence (conduction)-band offset at the interface, and $\Delta V$ the
    change of macroscopic-averaged electrostatic potential across the
    interface. Here, VBO (CBO) and $\Delta V$ are defined to have both positive
    and negative values; positive values of them indicate that the highest
    valence (lowest conduction) band or the macroscopic average of potential is shifted upward, crossing the interface from the NI region to the TI region. Negative values signify the opposite.}
  \ra{1.2}
    \begin{tabular}{ p{0.7in} p{0.3in} p{0.3in} p{0.4in} p{0.3in} p{0.35in}l  
                                           p{0.3in} p{0.3in} p{0.4in} p{0.3in} p{0.35in}l
                                           p{0.3in} p{0.3in} p{0.4in} p{0.3in} p{0.3in}}
    \hline
    \hline
    \multirow{2}{*}{eV units} &
    \multicolumn{5}{c}{Bi$_2$Se$_3(0001)$/InP(111)A} &&
    \multicolumn{5}{c}{Bi$_2$Se$_3$(0001)/InP(111)B} && 
    \multicolumn{5}{c}{Bi$_2$Te$_3$(0001)/BaF$_2$(111)} \\
     & $\Delta V$ & $E_\text{g}^\text{InP}$ & $E_\text{g}^\text{Bi$_2$Se$_3$}$ & VBO & CBO && $\Delta V$ & $E_\text{g}^\text{InP}$ & $E_\text{g}^\text{Bi$_2$Se$_3$}$ & VBO & CBO && $\Delta V$ & $E_\text{g}^\text{BaF$_2$}$ & $E_\text{g}^\text{Bi$_2$Te$_3$}$ & VBO & CBO \\
    \hline
    PBE+SO & \multirow{2}{*}{2.08} & 0.72 & 0.29 & 0.65 & 0.22 && \multirow{2}{*}{1.68} & 0.72 & 0.29 & 0.25 & -0.18 && \multirow{2}{*}{-1.52} & 6.82 & 0.08 & 3.19 & -3.55 \\
    $G_0W_0$+SO & & 1.42 & 0.21 & 1.01 & -0.21 && & 1.42 & 0.21 & 0.61 & -0.61 && & 9.98 & 0.16 & 5.27 & -4.55 \\
    \hline
    \hline
  \end{tabular}
  \label{tab:Tab2}
\end{table*}

Next, we turn our attention to the electronic structures of interface models
determined in Sec.~\ref{sec:3a} and first study band alignment at
their heterointerfaces. Figures~\ref{fig:Fig-macro}(a)
and~\ref{fig:Fig-macro}(b) display the planar and macroscopic averages of electrostatic potential along with the
ball-and-stick models of the corresponding heterojunction structures for the Bi$_2$Se$_3$(0001)/InP(111)A and
B interfaces, respectively. 
To obtain these potential plots, we performed the separate bulk and interface
calculations as described in
Sec.~\ref{sec:2}. Through bulk calculations for individual bulk components
of heterojunctions, we determined the reference level by use of the
macroscopic average method and the band-edge positions of the VBM and
CBM with respect to this reference level. Then, alignment of the each
reference line in bulk components was done through interface
calculations, yielding the potential lineup.
In bulk
calculations, we used the same structures as those used in building
interface components; that is, for bulk TIs, instead of their
experimental structures we used the structures whose in-plane lattice constants are set equal to those of the NI
substrates.
We also
used the experimental structures of bulk TIs\cite{Nakajima1963,Roy2014}, but due to the small lattice mismatch,
irrespective of the adopted structures, the difference in results is
negligible with the order of a few meV.
Because of the
polar nature of InP(111)A and B blocks in our heterojunction models, the macroscopic-averaged
values of potential in these systems don't tend to a constant in the bulk-like InP
region far from the interface, and instead they become a linearly
sloped potential in the bulk-like region, especially in the case of Bi$_2$Se$_3$(0001)/InP(111)A.
A close look at the plot in Fig.~\ref{fig:Fig-macro}(a) reveals that the
macroscopic-averaged potential is linearly ramped with the slope
of about 9~meV/\AA{}.
Thus, we evaluated the band offset by extending the linearly sloped
macroscopic-averaged potential in the bulk-like region to the midplane
between the interfacial layers and taking the value
on it as proposed in the previous literature\cite{Picozzi1997,Amico2015}.

Through interface calculations we found that electrostatic potential shifts upward by 2.08~eV and 1.68~eV, as it traverses the
interface from the InP to the Bi$_2$Se$_3$ region, for the
Bi$_2$Se$_3$(0001)/InP(111)A and B heterojunctions, respectively. Bulk
DFT-PBE calculations give the values of 0.29~eV
and 0.72~eV for the band gaps of Bi$_2$Se$_3$ and InP, respectively.
These band gaps are different from the corresponding experimental values due to the
well-known limitation of DFT for the prediction of band gaps and these
deviations are transferred to the conduction-band offsets. 
As for the valence (conduction)-band offsets, from semilocal
DFT-PBE calculations we obtained the values of
0.65~eV (0.22~eV) and 0.25~eV (0.18~eV) for interface A and B, respectively.
Then, we corrected the
band gap and band offset by using the $G_0W_0$-corrected band gap
and band-edge positions. For instance, the $G_0W_0$-corrected
valence (conduction)-band offset $E_\text{v}^\text{$G_0W_0$,BO}$ ($E_\text{c}^\text{$G_0W_0$,BO}$) is calculated as follows:
\begin{eqnarray}
  \label{eq:5}
  E_\text{v}^\text{$G_0W_0$,BO} &=& E_\text{v}^\text{BO}+\Delta
                             E_\text{VBM}^\text{$G_0W_0$}\\
  E_\text{c}^\text{$G_0W_0$,BO} &=& E_\text{c}^\text{BO}+\Delta
                             E_\text{CBM}^\text{$G_0W_0$}+\Delta E_\text{g}^\text{$G_0W_0$}\,,
\end{eqnarray}
where $E_\text{v(c)}^\text{BO}$ is the band offset at the level of
DFT-PBE, $\Delta E_\text{VBM(CBM)}^\text{$G_0W_0$}$ is the
$G_0W_0$-corrected band-edge position for the valence (conduction) band, and $\Delta E_\text{g}^\text{$G_0W_0$}$ is the $G_0W_0$-corrected band
gap.

\begin{figure}[hb!]
  \centering
  \includegraphics[width=0.5\textwidth]{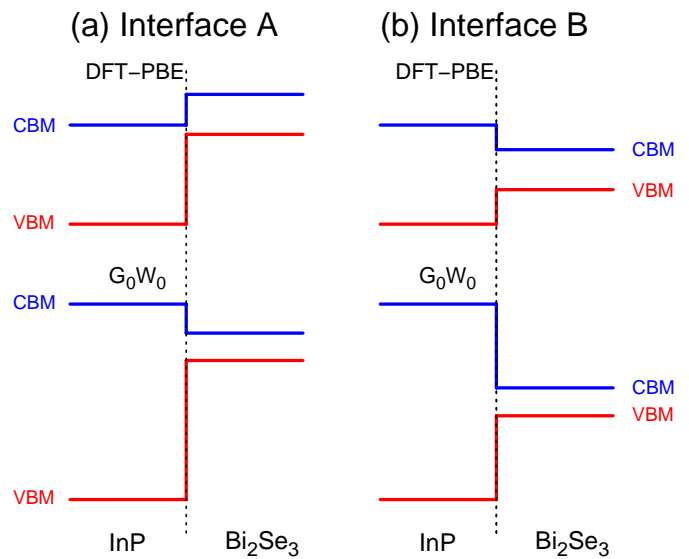}
  \caption{(Color online) Schematic plot of the band
      alignment for (a) Bi$_2$Se$_3$(0001)/InP(111)A (Interface
    A) and (b) Bi$_2$Se$_3$(0001)/InP(111)B (Interface B). Here, the
    band diagrams are shown, which are obtained within both the levels of
    DFT-PBE and $G_0W_0$. The band-edge positions of VBM and CBM are
    indicated by the red-colored and blue-colored solid lines,
    respectively.
    Black-dashed lines correspond to the interface.}
  \label{fig:Fig-diagram}
\end{figure}

Performing $G_0W_0$ calculations changes these values to, for
example, 0.21~eV and 1.42~eV for a bulk band gap of
Bi$_2$Se$_3$ and InP, respectively.
These calculated band gaps are in good agreement with the experimental ones and
especially, our band gaps of bulk Bi$_2$Se$_3$ and Bi$_2$Te$_3$ agree well with those
from optical experiments\cite{Orlita2015,Thomas1992} and those obtained
from the full-potential calculation with the same treatment of
spin-orbit coupling\cite{Aguilera2013}.
Putting together, we found that $G_0W_0$ calculations change the
valence (conduction)-band offset to 1.01~eV (0.21~eV) for interface A and
0.61~eV (0.61~eV) for interface B.

For the Bi$_2$Te$_3$(0001)/BaF$_2$(111) heterojunction,
our interface calculations have shown that electrostatic
  potential shifts downward by 1.52~eV, crossing the interface from the
  BaF$_2$ to the Bi$_2$Te$_3$ region, and bulk DFT-PBE calculations give the band gaps of
0.08~eV and 6.82~eV for Bi$_2$Te$_3$ and BaF$_2$, respectively, which increase
to 0.16~eV and 9.98~eV in $G_0W_0$ calculations.
For the valence (conduction)-band offset, we obtained 3.19~eV
(3.55~eV) from semilocal
DFT-PBE calculations which changes to 5.27~eV (4.55~eV) in $G_0W_0$ calculations.

All results are collected in Table~\ref{tab:Tab2} and the
schematic band-alignment diagrams for the two types of Bi$_2$Se$_3$(0001)/InP(111) heterojunctions are shown in Fig.~\ref{fig:Fig-diagram}.
As can be seen from this figure, within the DFT-PBE level interface A
and B show the type-II and type-I band alignment, respectively, however within the
$G_0W_0$ level, both of them exhibit the type-I band alignment.

\subsubsection{\label{sec:3b2}Topological interface states}

Finally, we focus on the electronic and spin structures
of the topological interface states in heterojunctions
considered. All results in this subsection are
  calculated within the DFT-PBE level. First, we identify the
interface states by projecting the
Kohn-Sham wave
functions onto a set of localized functions defined in spheres centered at the atomic sites of
the layers around the interface. 
The radius of sphere is set to one-half of the nearest-neighbor
distance in the system and the localized function is modeled with a constant radial part
multiplied by real spherical
harmonics and normalized to one.

\begin{figure}
  \centering
  \includegraphics[width=0.475\textwidth]{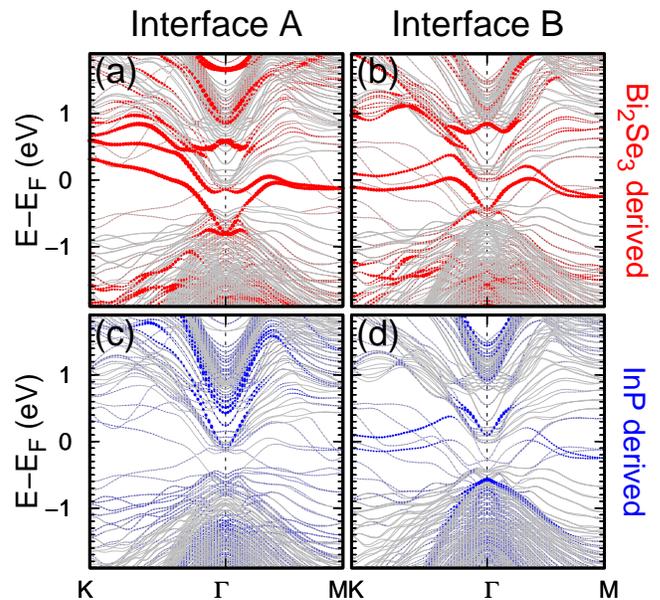}
  \caption{(Color online) Band structures along high-symmetry $\bm{k}$ directions decorated with the variable-sized dots proportional to the weights of projection
    of wave functions onto the near-interface region for (a,c) Bi$_2$Se$_3$(0001)/InP(111)A (Interface
    A) and (b,d) Bi$_2$Se$_3$(0001)/InP(111)B (Interface B). Red- and
    blue-colored dots represent the interface states derived from
    Bi$_2$Se$_3$ and InP atomic layers around the interface,
    respectively.}
  \label{fig:proj-BSIP}
\end{figure}

Figure~\ref{fig:proj-BSIP} presents the results for two types of
Bi$_2$Se$_3$(0001)/InP(111) heterojunctions. Here, the 1-QL
Bi$_2$Se$_3$ or the 6-layer InP closest to the interface
is taken as the projection region to identify the interface states
confined to within the corresponding region and the size of filled
circles is proportional to the projection weight.
As depicted in Figs.~\ref{fig:proj-BSIP}(a) and~\ref{fig:proj-BSIP}(b),
topological interface states are clearly visible near the $\Gamma$
point and they are mainly derived from the Bi$_2$Se$_3$ side of the
interface, which is in line with our expectation. In particular, the
Dirac-cone character is more prominent in Bi$_2$Se$_3$(0001)/InP(111)A (Interface A) than in
Bi$_2$Se$_3$(0001)/InP(111)B (Interface B), which can be attributed to the fact that in
the latter InP-derived interface states are also formed near the Dirac
point (DP) as seen in Fig.~\ref{fig:proj-BSIP}(d) and possibly hybridized with topological interface states.

Additionally, in our heterojunction models we can see the feature bearing on the
  vertical twinning of DPs as reported in the previous
  study\cite{Seixas2015} although the upper Dirac cone is less
  clear than the lower one. We also find that the location of DPs is
  dependent on interface details. 
  Namely, in
  interface A the lower DP is located around 0.4~eV below the
  VBM, whereas it resides nearly at the VBM in interface B. This difference can be explained by the difference of potential lineup between
interface A and B, and is consistent with the fact that
interface states are subject to the band-bending effect in
heterojunctions since they are usually localized within the region where
potential changes abruptly (see Fig.~\ref{fig:Fig-macro})\cite{Menshov2015,Seixas2015}.

\begin{figure}
  \centering
  \includegraphics[width=0.45\textwidth]{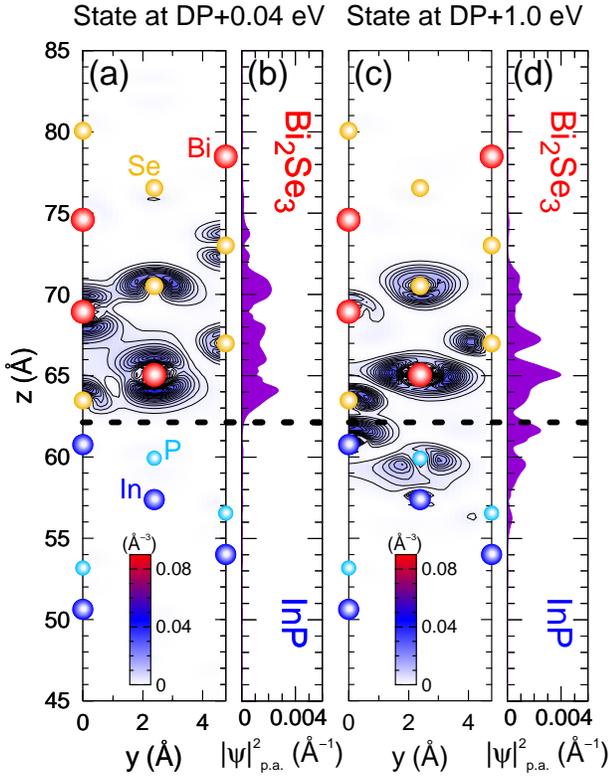}
  \caption{(Color online) Plot of squared wave functions of the
    topological interface states for Bi$_2$Se$_3$(0001)/InP(111)A at the
    energy of (a,b) 0.04~eV and (c,d) 1.0~eV above the DP. (a) and (c) present the color-shaded contour plot in the $x=0$ plane, and
    (b) and (d) the one-dimensional planar-averaged (p.a.) plot of (a) and
    (c), respectively. The interval between adjacent contour lines is
    2.5$\times$10$^{-3}$ per \AA$^3$.
    The atoms near the interface are marked by the balls with the corresponding colors and some of
    them are denoted with the labels of the same color.
  }
  \label{fig:wfn}
\end{figure}

Second, we examine the degree of localization of the topological interface
states. Figure~\ref{fig:wfn} shows the spatial distribution of
squared wave functions of topological interface states for
Bi$_2$Se$_3$(0001)/InP(111)A. As we can see in Figs.~\ref{fig:wfn}(a) and~\ref{fig:wfn}(b), the topological interface states near
the DP are
strictly confined to within about~12~\AA{} around the interface and they
hardly extend into the InP region. As the energy of the interface
states goes away from the DP, however, they appear to start to penetrate
into the InP side. For instance, the wave function of the interface state at the energy of
1.0~eV above the DP enters into the region of InP, with the
third subinterface P layer having about one-tenth of its maximum
density [see Figs.~\ref{fig:wfn}(c) and~\ref{fig:wfn}(d)].

Lastly, we investigate the spin structures of topological interface
states. As already mentioned, the helical spin textures of
topologically protected boundary states make a TI distinguishable from the integer quantum Hall system and are relevant for their
application to, in particular, spintronics in which long spin
coherence is crucial\cite{Wolf2001}. Along this line, we obtained the spin textures of topological interface
states for the Bi$_2$Se$_3$(0001)/InP(111)A heterojunction by calculating the expectation values of the Pauli spin operator,
$\sigma_i$ ($i=x,y,z$), for them. 

\begin{figure}[t!]
  \centering
  \includegraphics[width=0.443\textwidth]{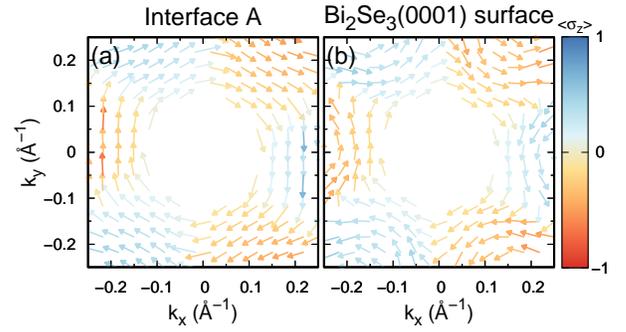}
  \caption{(Color online) Spin textures of (a) topological interface states for Bi$_2$Se$_3$(0001)/InP(111)A (Interface A) and (b) topological surface
    states at the vacuum-facing surface of 6-QL Bi$_2$Se$_3$ slab which is
    constructed from Bi$_2$Se$_3$(0001)/InP(111)A by
    removing the part of InP. In both
  cases, only states above the DP are considered.
Additionally, the spin
textures near the zone center are excluded since in this region the topological interface states for
Bi$_2$Se$_3$(0001)/InP(111)A are buried under the bulk-like states
[see Fig.~\ref{fig:proj-BSIP}(a)] and
thus their spin textures are not well resolved.
 The length
  and direction of arrows indicate, respectively, the magnitude and direction
  of the expectation values of the in-plane ($\sigma_x$ and
  $\sigma_y$) components of the Pauli spin operator, while their color represents the
  normal-to-surface component ($\sigma_z$).}
  \label{fig:spin}
\end{figure}

Figures~\ref{fig:spin}(a) and~\ref{fig:spin}(b) present these
results for Bi$_2$Se$_3$(0001)/InP(111)A and 6-QL Bi$_2$Se$_3$, respectively.
From them, we can see that spin textures exhibit the similar tendency
of a nearly circular form and a clockwise direction for the states
above the DP, which is well in accord with the previous results for those at the vacuum-facing
surfaces\cite{Yazyev2010,Park2012}. The only difference is that in the considered
momentum range, the topological interface states for
Bi$_2$Se$_3$(0001)/InP(111)A show the less warped helical spin
texture\cite{Fu2009} as compared with the topological surface states at the vacuum-facing
surfaces\cite{Kuroda2010}. It can be attributed to the lowering of DP
and the smaller Fermi
velocity of 3.10~eV$\cdot$\AA{} than the experimentally measured value
of 3.55~eV$\cdot$\AA{} of the Dirac-cone dispersion on the
vacuum-facing surfaces\cite{Kuroda2010}, which in turn induce less
hybridization with the bulk states.
Except for this moderate difference, our results indicate that the spin-helical structures of topological interface states are as well maintained as those at the vacuum-facing
surfaces.

For the Bi$_2$Te$_3$(0001)/BaF$_2$(111) heterojunction, all tendencies above are
similar and because of the much larger bulk band gap of BaF$_2$, most properties of
topological interface states are indiscernible from those at
the vacuum-facing surfaces.

\section{\label{sec:4}Conclusion}

In summary, based on quasiparticle $GW$ approximation as well as
semilocal DFT we have presented the results of a theoretical study of
the experimentally realized
lattice-matched heterojunctions, Bi$_2$Se$_3$(0001) on InP(111) and
Bi$_2$Te$_3$(0001) on BaF$_2$(111), focusing on the band offsets at these heterointerfaces and
the electronic structures and spin
textures of the topological interface states. 
Topological interface states are shown to be strictly localized at these lattice-matched
heterointerfaces. We further demonstrated that their helical spin textures are as well maintained as those at the
vacuum-facing surfaces. Along with these similarities, topological interface states also exhibit some
differences from topological surface states in the extent of hexagonal
warping of spin textures and lowering of the DP, both of which can be ascribed to the
band-bending effect occurring at the heterointerfaces of these systems.
Taken collectively, our results point to the
potential uses of these lattice-matched TI/NI heterojunctions in the future
spintronic and electronic solid-state devices that build on
topological insulators.

\begin{acknowledgments}
  We were supported by the
  European Research Council starting grant ``TopoMat'' (Grant
  No. 306504). First-principles computations have been performed at
  the Swiss National Supercomputing Centre under Project Nos. s515 and s675.
\end{acknowledgments}

\bibliography{reference}

\end{document}